
\def\J{$J/\psi$}
\def\j{J/\psi}

\def\p{\psi'}

\def\c{c{\bar c}}

\def\q{q{\bar q}}

\def\lsim{\raise0.3ex\hbox{$<$\kern-0.75em\raise-1.1ex\hbox{$\sim$}}}
\def\gsim{\raise0.3ex\hbox{$>$\kern-0.75em\raise-1.1ex\hbox{$\sim$}}}

\newcount\REFERENCENUMBER\REFERENCENUMBER=0
\def\REF#1{\expandafter\ifx\csname RF#1\endcsname\relax
               \global\advance\REFERENCENUMBER by 1
               \expandafter\xdef\csname RF#1\endcsname
                   {\the\REFERENCENUMBER}\fi}
\def\reftag#1{\expandafter\ifx\csname RF#1\endcsname\relax
               \global\advance\REFERENCENUMBER by 1
               \expandafter\xdef\csname RF#1\endcsname
                      {\the\REFERENCENUMBER}\fi
             \csname RF#1\endcsname\relax}
\def\ref#1{\expandafter\ifx\csname RF#1\endcsname\relax
               \global\advance\REFERENCENUMBER by 1
               \expandafter\xdef\csname RF#1\endcsname
                      {\the\REFERENCENUMBER}\fi
             [\csname RF#1\endcsname]\relax}
\def\refto#1#2{\expandafter\ifx\csname RF#1\endcsname\relax
               \global\advance\REFERENCENUMBER by 1
               \expandafter\xdef\csname RF#1\endcsname
                      {\the\REFERENCENUMBER}\fi
           \expandafter\ifx\csname RF#2\endcsname\relax
               \global\advance\REFERENCENUMBER by 1
               \expandafter\xdef\csname RF#2\endcsname
                      {\the\REFERENCENUMBER}\fi
             [\csname RF#1\endcsname--\csname RF#2\endcsname]\relax}
\def\refs#1#2{\expandafter\ifx\csname RF#1\endcsname\relax
               \global\advance\REFERENCENUMBER by 1
               \expandafter\xdef\csname RF#1\endcsname
                      {\the\REFERENCENUMBER}\fi
           \expandafter\ifx\csname RF#2\endcsname\relax
               \global\advance\REFERENCENUMBER by 1
               \expandafter\xdef\csname RF#2\endcsname
                      {\the\REFERENCENUMBER}\fi
            [\csname RF#1\endcsname,\csname RF#2\endcsname]\relax}
\def\refss#1#2#3{\expandafter\ifx\csname RF#1\endcsname\relax
               \global\advance\REFERENCENUMBER by 1
               \expandafter\xdef\csname RF#1\endcsname
                      {\the\REFERENCENUMBER}\fi
           \expandafter\ifx\csname RF#2\endcsname\relax
               \global\advance\REFERENCENUMBER by 1
               \expandafter\xdef\csname RF#2\endcsname
                      {\the\REFERENCENUMBER}\fi
           \expandafter\ifx\csname RF#3\endcsname\relax
               \global\advance\REFERENCENUMBER by 1
               \expandafter\xdef\csname RF#3\endcsname
                      {\the\REFERENCENUMBER}\fi
[\csname RF#1\endcsname,\csname RF#2\endcsname,\csname
RF#3\endcsname]\relax}
\def\refand#1#2{\expandafter\ifx\csname RF#1\endcsname\relax
               \global\advance\REFERENCENUMBER by 1
               \expandafter\xdef\csname RF#1\endcsname
                      {\the\REFERENCENUMBER}\fi
           \expandafter\ifx\csname RF#2\endcsname\relax
               \global\advance\REFERENCENUMBER by 1
               \expandafter\xdef\csname RF#2\endcsname
                      {\the\REFERENCENUMBER}\fi
            [\csname RF#1\endcsname,\csname RF#2\endcsname]\relax}
\def\Ref#1{\expandafter\ifx\csname RF#1\endcsname\relax
               \global\advance\REFERENCENUMBER by 1
               \expandafter\xdef\csname RF#1\endcsname
                      {\the\REFERENCENUMBER}\fi
             [\csname RF#1\endcsname]\relax}
\def\Refto#1#2{\expandafter\ifx\csname RF#1\endcsname\relax
               \global\advance\REFERENCENUMBER by 1
               \expandafter\xdef\csname RF#1\endcsname
                      {\the\REFERENCENUMBER}\fi
           \expandafter\ifx\csname RF#2\endcsname\relax
               \global\advance\REFERENCENUMBER by 1
               \expandafter\xdef\csname RF#2\endcsname
                      {\the\REFERENCENUMBER}\fi
            [\csname RF#1\endcsname--\csname RF#2\endcsname]\relax}
\def\Refand#1#2{\expandafter\ifx\csname RF#1\endcsname\relax
               \global\advance\REFERENCENUMBER by 1
               \expandafter\xdef\csname RF#1\endcsname
                      {\the\REFERENCENUMBER}\fi
           \expandafter\ifx\csname RF#2\endcsname\relax
               \global\advance\REFERENCENUMBER by 1
               \expandafter\xdef\csname RF#2\endcsname
                      {\the\REFERENCENUMBER}\fi
        [\csname RF#1\endcsname,\csname RF#2\endcsname]\relax}
\def\refadd#1{\expandafter\ifx\csname RF#1\endcsname\relax
               \global\advance\REFERENCENUMBER by 1
               \expandafter\xdef\csname RF#1\endcsname
                      {\the\REFERENCENUMBER}\fi \relax}

%

\def\NP{{ Nucl.\ Phys.\ }}
\def\PL{{ Phys.\ Lett.\ }}
\def\PR{{ Phys.\ Rev.\ }}

\def\PRL{{ Phys.\ Rev.\ Lett.\ }}

\def\q{q{\bar q}}

\magnification=1200
\hsize=16.0truecm
\vsize=24.5truecm
\baselineskip=13pt
\def\q{\q{\bar q}}

\pageno=0
{}~~~
\hfill CERN-TH/95-27\par
\hfill BI-TP 95/15\par
\hfill TPI-MINN 95/07\par
\hfill TH 1336
\vskip 2.5truecm
\centerline{\bf NON-PERTURBATIVE QUARKONIUM DISSOCIATION}
\medskip
\centerline{\bf IN HADRONIC MATTER}
\vskip 1 truecm
\centerline{D. Kharzeev$^a$, L. McLerran$^b$ and H. Satz$^a$}
\bigskip \bigskip
\centerline{$^a$ Theory Division, CERN, CH-1211 Geneva, Switzerland}
\centerline{and}
\centerline{Fakult\"at f\"ur Physik, Universit\"at Bielefeld,
D-33501 Bielefeld, Germany}
\bigskip
\centerline{$^b$ School of Physics and Astronomy, University of Minnesota,
MN 55455 Minneapolis, USA}
\vskip 2 truecm
\centerline{\bf Abstract}
\medskip
We calculate the dissociation rates of quarkonium ground states by
tunnelling and direct thermal activation to the continuum. For
hadronic matter at temperatures $T \leq 0.2$ GeV, neither of
these mechanisms leads to a sufficiently large dissociation
to explain the experimentally observed suppression of charmonium.
Dissociation by sequential excitation to excited energy levels,
although OZI-forbidden, requires further analysis.

\par\vfill\noindent
CERN-TH/95-27\par\noindent
BI-TP 95/15\par\noindent
TPI-MINN 95/07\par\noindent
TH 1336\par\noindent
April 1995
\eject

\def\p{\psi}
\def\la{\Lambda_{\rm QCD}}
\bigskip\bigskip

\bigskip \bigskip
\refadd{Peskin}
\refadd{Vain}
\refadd{Kaida}
\refadd{Quarko}
For sufficiently heavy quarks, the dissociation of quarkonium states
by interaction with light hadrons can be fully accounted for
by short-distance QCD \refto{Peskin}{Quarko}. Perturbative calculations
become valid
when the space and time scales associated with the quarkonium state,
$r_Q$ and $t_Q$, are small in comparison to the nonperturbative
scale $\la^{-1}$
$$
r_Q<<\la^{-1}, \eqno(1a)
$$
$$
t_Q<<\la^{-1}; \eqno(1b)
$$
$\la^{-1}$ is also the characteristic size of the light hadrons.
In the heavy quark limit, the quarkonium binding becomes Coulombic,
and the spatial size $r_Q \sim (\alpha_sm_Q)^{-1}$ thus is small.
The time scale is by the uncertainty relation given as the inverse
of the binding energy $E_Q \sim m_Q$ and hence also small.
For the charmonium ground state $\j$, we have
$$
r_{\p} \simeq 0.2~{\rm fm}= (1~ {\rm GeV})^{-1} \eqno(2)
$$
and
$$
E_{\p} = 2 M_{D} - M_{\p} \simeq 0.64~{\rm GeV}. \eqno(3)
$$
With $\la \simeq 0.2$ GeV, the inequalities (1) seem already
reasonably well satisfied, and also the heavy quark relation
$E_{\p} = (1/m_c r_{\p}^2)$ is very well fulfilled.
We therefore expect that the dissociation of \J's in hadronic matter
will be governed by the \J-hadron break-up cross
section as calculated in short-distance QCD.
\par
Nevertheless, in view of the finite charm quark mass, it makes sense
to study possible non-perturbative contributions, in particular for the
calculation of the break-up process in matter, where in addition
thermal activation may come into play.
For an isolated \J-hadron system, non-perturbative interactions can be
pictured most simply as a quark rearrangment. Consider
putting a \J~``into" a stationary light hadron;
the quarks could then just rearrange their binding pattern
to give rise to transitions such as $\j + N \to \Lambda_c+{\bar D}$ or
$\j+\rho \to D + {\bar D}$ (Fig.\ 1). The probability for such a
transition can be written as
$$
P_{\rm rearr} \sim \int d^3 r~R(r)~|\phi_{\psi}(r)|^2, \eqno(4)
$$
where the spatial distribution of the $\c$ bound state is
given by the squared wave function $|\phi_{\psi}(r)|^2$. The function
$R(r)$ in eq.\ (4) describes the resolution capability of the colour
field inside the light hadron. Its wave length is of order $\la^{-1}$,
and so it cannot resolve the charge content of very much smaller bound
states; in other words, it does not ``see" the heavy quarks in a
bound state of
radius $r_Q<< \la^{-1}$ and hence cannot rearrange bonds.
The resolution $R(r)$ will approach unity for $r\la >>1$
and drop very rapidly with $r$ for $r\la < 1$, in the
functional form
$$
R(r) \simeq (r\la)^n, ~~r\la < 1,   \eqno(5)
$$
with $n=2$ \ref{Low} or 3 \ref{Peskin}.
As a result, the integrand of eq.\ (4)
will peak at some distance $r_0$, with $r_Q< r_0 < \la$.
Since the bound state radius of the
quarkonium ground state decreases with increasing heavy quark mass,
while $R(r)$ is $m_Q$-independent, $r_0 \to r_Q \to 0$ as $m_Q \to
\infty$. Hence $P_r$ vanishes in the limit $m_Q \to \infty$
because $R(r_0)$ does, indicating that
the light quarks can no longer resolve the small heavy quark
bound state.
\par
In a potential picture, the situation just described means that the
charm quarks inside the
\J~have to tunnel from $r=r_{\psi}$ out to a distance at which the
light quarks can resolve them, i.e., out to some $r\simeq c \la^{-1}$,
where $c$ is a constant of order unity (Fig.\ 2).
Such tunneling processes are therefore
truly non-perturbative: they cover a large space-time
region, of linear size $\Lambda_{QCD}^{-1}$, and do not involve any
hard interactions. The first aim of this letter is to estimate
the contribution of non-perturbative tunnelling to the dissociation
of quarkonium states. Following this,
we turn to matter at finite temperature, where the \J~can be excited
into the continuum by thermal activation. Our second objective is
to calculate the rate of dissociation by this mechanism.
\par
In general, the problem of quark tunneling cannot be solved in a
rigorous way, since it involves genuine non-perturbative QCD
dynamics. However, the large mass of the heavy quark allows a
very important simplification, the use of the quasiclassical
approximation. In this approximation, the rate of tunneling
$R_{\rm tun}$ can be written down in a particularly transparent way:
it is simply the product of the frequency $\omega_{\p}$ of the heavy
quark motion in the potential well and the tunnelling probability
$P_{\rm tun}$ when the quark hits the wall of the well,
$$
R_{\rm tun} =\omega_{\p} P_{\rm tun} \eqno(6)
$$
The frequency $\omega_{\p}$ is
determined by the gap to the first radial excitation,
$$
\omega_{\p} \simeq (M_{\p'}-M_{\p}) \simeq \ E_{\p}. \eqno(7)
$$
Consider now the potential seen by the $\c$ (Fig.\ 2).
For a particle of energy $E$,
the probability of tunneling through the potential barrier $V(r)$
is obtained from the squared wave function in the ``forbidden" region.
It can be expressed in terms of the action $W$ calculated along the
quasiclassical trajectory,
$$
P_{\rm tun} = e^{-2W}, \eqno(8)
$$
where
$$
W= \int_{r_1}^{r_2} |p|\ dr, \eqno(9)
$$
Here the momentum $|p|$ is given by
$$
|p|=\left[ 2M(V(x)-E) \right]^{1/2}, \eqno(10)
$$
and $r_1, r_2$ are the turning points of the classical motion
determined from the condition $ V(r_i)=E $.
\par
In our case, the width of the barrier is approximately $0.6~\la^{-1}$,
while its height $(V - E)$ is equal to the dissociation threshold
$E_Q$. The mass $M$ in Eq.(10) is the reduced mass, $M=m_Q/2$.
We thus have
$$
W \simeq 0.6~\sqrt{m_Q\ E_Q}/\la. \eqno(11)
$$
For the \J, we get from (11) the value $W \simeq 3$; this
$a\ posteriori$ justifies the use of quasiclassical approximation,
which requires $S>1$.

Using eq.\ (11), we obtain as final form for the tunneling rate
(6)
$$
R_{\rm tun} = E_{\p} \
{\rm exp} - (1.2 \sqrt{m_c\ E_{\p}}/ \la).
\eqno(12)
$$
With the above mentioned \J~parameters, this leads to the
very small dissociation rate
$$
R_{\rm tun} \simeq 9.0 \times 10^{-3}~{\rm fm}^{-1}.\eqno(13)
$$
In terms of $R_0$, the \J~survival probability is given by
$$
S_{\rm tun} = {\rm exp}\left\{
-\int_0^{t_{\rm max}} dt~R_{\rm tun}\right\}, \eqno(14)
$$
where $t_{\rm max}$ denotes the
maximum time the \J~spends adjacent to the light hadron. In the
limit $t_{\rm max} \to \infty$, $S_{\p}$ vanishes. However,
the uncertainty relations prevent a localisation of the two systems
in the same spatial area for long times. From $\Delta x \leq
\la^{-1}$ we get $\Delta p \geq \la$, so
that the longest time which the \J~can spend in the interaction
range of the light hadron is
$$
t_{\rm max} = \la^{-1} \left( 1 + {m^2 \over
\la^2} \right)^{1/2}, \eqno(15)
$$
with $m$ for the mass of the light hadron. For nucleons or vector
mesons, this time is 4 - 5 fm, and with this, the survival probability
is very close to unity, so that tunnelling cannot result
in a noticeable non-perturbative \J~dissociation.
\par
In a medium at finite temperature, however, the \J~can in addition
be thermally excited into the continuum and thus become dissociated;
this thermal activation could be non-perturbative. Here we shall simply
consider a \J~in a thermal medium of temperature $T$ and calculate its
excitation rate, without asking how the constituents
of this medium bring the ground state $\c$ into the continuum.
Since hadrons in a medium of temperature $T$ may not be able to interact
sufficiently hard with the \J~to overcome the mass gap to the continuum,
we obtain in this way an upper bound to direct thermal dissociation.
\par
The partition function of the system at finite temperature is
$$
Z(T)=\sum_n e^{-E_n/T} = Z_{\p}(T) + Z_{\rm cont}(T), \eqno(16)
$$
where $Z_{\rm cont}$ and $Z_{\p}$ are the continuum and the bound-state
contributions, respectively. We assume that the $\c$ will be distributed
among the ground state \J~and the continuum above $E=2M_D$ according to
$$
\rho(E,T) = c(E)~e^{-E/T}[\delta(1-E/M_{\p})
+ \Theta(E-2M_D)], \eqno(17)
$$
where $c(E)$ denotes the degeneracy factor, which we take to be
constant, $c(E)=c$. The continuum part of the partition function is
given by
$$
Z_{cont}(T) = V c \int {d^3p \over (2\pi)^3} e^{-E/T} \Theta(E-2M_D),
\eqno(18)
$$
where $V$ is the canonical volume containing the system. Changing
variables to
$$
 p^2 dp = M \sqrt{2ME} dE, \eqno(19)
$$
we get
$$
Z_{\rm cont}(T) = V \bar{c}\ {4\pi \over (2\pi)^3}\ M\ \sqrt{\pi M \over 2}\
T^{3/2}\ e^{-E_{\p}/T}, \eqno(20)
$$
while the bound-state part is obtained from eq.\ (17) as
$$
Z_{\p}(T) =  \bar{c}, \eqno(21)
$$
with $E_{\p}= 2M_D-M_{\p}$ and ${\bar c}\equiv c~e^{-M_{\p}/T}$.
\par
The rate of escape into the continuum can be estimated by the average
time needed to leave the spatial region of the potential well,
$$
R_{\rm act} = \langle t \rangle^{-1} = \langle v(E)\rangle /L.
\eqno(22)
$$
Here $v(E)$ is the velocity of the (reduced) charm quark in the
continuum, and $L\simeq (1 - r_{\p}\la)~\la^{-1}$ is the distance from
the average \J~radius to the top to the potential well. Changing
variables from the overall energy to the energy above $M_{\p}$ and
using $v(E)=p(E)/M$, we bring eq.\ (22) into the form
$$
R_{\rm act} = {1\over Z(T)}{V\over{ML}}
\int {d^3p\over{(2\pi)^3}}\ p\ [\bar{c}\ \Theta(E-E_{\p}) e^{-E/T}].
\eqno(23)
$$
With the substitution $p(E)=\sqrt{2M(E-E_{\p})}$ we then have
after integration the result
$$
R_{\rm act} = {1\over Z(T)}{V\over{L}} \left({\bar{c}\over{\pi^2}} MT^2 \right)
e^{-E_{\p}/T},
\eqno(24)
$$
where $V=L^3$ is the canonical volume.
\par
For high temperatures, $T\gg E_{\p}$, the continuum gives the dominant
contribution to the statistical sum, so that we can replace
$Z(T)$ by $Z_{\rm cont}$ in eq.\ (24) to get
$$
R_{\rm act}= {4 \over L} \sqrt{ T \over{2\pi M}}. \eqno(25)
$$
Recalling that the thermal velocity of a free particle in three
dimensions is just $v_{\rm th}(T)=4 \sqrt{ T / 2\pi M}$, we thus recover
the classical high-temperature limit for the thermal activation rate
$$
R_{\rm act}= {v_{\rm th}(T) \over L}. \eqno(26)
$$
At low temperatures, for $T\ll E_{\p}$, the discrete term in
$\rho$ give the main contribution to $Z(T)$. In this temperature
range, which is the one of interest here, we thus obtain from eqs.\
(24) and (21)
$$
R_{\rm act}= {(LT)^2 \over {3\pi}}\ M\
e^{-E_{\p}/T},  \eqno(27)
$$
as our final result.
\par
The explicit appearance of the volume factors $L$ in the above results
for thermal activation may at first sight seem strange. Its origin is
the fact that although the intrinsic Boltzman factor for excitations
to the continuum is small, the density of states there is quite large.
Nevertheless, the limit of large $L$ is well defined,
since for $L\to \infty$ the continuum part (20) of the
partition function dominates, so that the result (27) is replaced by
the classical formula (25). The heavy quark limit $M \to \infty$ is also
well defined, since at very large $M$ the size of the system shrinks
to $L\sim 1/M$. It is moreover important to remember that for heavy
quark-antiquark systems the binding energy (on which the rate (27)
depends exponentially) increases with the mass of the quark.
\par
With the same charmonium parameters as above ($E_{\p}=0.64$ GeV, $L=0.6
{}~\la^{-1}$, $2M=m_c=1.5$ GeV) and at the temperature $T=0.2$ GeV, we
thus obtain
$$
R_{\rm act} \simeq 6 \times 10^{-3}~{\rm fm}^{-1},\eqno(28)
$$
which is of the same size as the tunnelling rate (13).
The precise value of $L$ is somewhat uncertain, of course. We
feel, however, that $L\simeq 1$ fm is an upper bound, leading to
$R_{\rm act} \leq 2 \times 10^{-2}\ {\rm fm}^{-1}$ as upper bound for
the rate. Thus direct thermal activation can also not provide a
significant amount of \J~dissociation at temperatures below the
binding energy. Already for static media of 4-5 fm life-time
the survival probability is very close to unity; any expansion would
further reduce activation effects.
\par
We should however be cautious about the application of the thermal
activation formula used above.  Because of the large density of states
in the continuum, the treatment of the factor $L$ was somewhat heuristic.
To better understand this dependence, it will probably be necessary to
investigate directly the sequential transitions from the ground
state to excited states and then to the continuum.
Thus the \J~could be dissociated by first
exciting it to a $\chi_c$ and then bringing the $\chi_c$ to the
continuum. Since the phase space for the continuum is so large, the
probability of de-excitation back to a bound state is negligible.
Unfortunately, it is difficult to obtain a simple estimate of
the various processes which contribute to this sequential excitation.
The analysis of this process awaits further work.
\bigskip
\par
D. K. and H. S. acknowledge the financial support of the
German Research Ministry (BMFT) under the Contract 06 BI 721.
The work of L. M. was supported in part by the U.S. Department
of Energy under Grants DOE High Energy DE-AC02-83ER40105 and
DOE Nuclear DE-FG02-87ER40328.

\vfill\eject

\centerline{\bf References}
\medskip
\item{\reftag{Peskin})}{M. E. Peskin, \NP B156 (1979) 365;\hfill\break
                    G. Bhanot and M. E. Peskin, \NP B156 (1979) 391.}
\item{\reftag{Vain})}{M. A. Shifman, A. I. Vainshtein and V. I.
Zakharov,
                    \PL 65B (1976) 255;\hfill\break
                    V. A. Novikov, M. A. Shifman, A. I. Vainshtein and
                     V. I. Zakharov, \NP B136 (1978) 125.}
\item{\reftag{Kaida})}{A. Kaidalov, in {\it QCD and High Energy
                    Hadronic Interactions}, J. Tr\^an Thanh V\^an (Ed.),
                    Edition Frontieres, Gif-sur-Yvette, 1993.}
\item{\reftag{Quarko})}{D. Kharzeev and H. Satz, \PL B 334 (1994) 155.}
\item{\reftag{Low})}{F. E. Low, \PR D12 (1975) 163;\hfill\break
                    S. Nussinov, \PRL 34 (1975) 1268;\hfill\break
                    J. F. Gunion and H. Soper, \PR D15 (1977) 2617.}

\vfill\eject

\centerline{\bf Figure Captions}
\bigskip\bigskip
{\bf Fig. 1}  Rearrangement transition $\rho + J/\Psi \to D + \bar{D}$.
\par
\medskip
{\bf Fig. 2}  $J/\Psi$ dissociation by tunneling.

\vfill\bye